# The State of the Heliosphere Revealed by Limb Halo Coronal Mass Ejections in Solar Cycles 23 and 24

Nat Gopalswamy[1], Sachiko Akiyama[1,2], and Seiji Yashiro[1,2]

[1]NASA Goddard Space Flight Center, Greenbelt, MD 20771

[2]The Catholic University of America, Washington DC 20624

Abstract

We compare the properties of halo coronal mass ejections (CMEs) that originate close to the limb (within a central meridian distance range of 60º to 90º) during solar cycles 23 and 24 to quantify the effect of the heliospheric state on CME properties. There are 44 and 38 limb halos in the cycles 23 and 24, respectively. Normalized to the cycle-averaged total sunspot number, there are 42% more limb halos in cycle 24. Although the limb halos as a population is very fast (average speed ~1464 km $s^{-1}$), cycle-24 halos are slower by ~26% than the cycle-23 halos. We introduce a new parameter, the heliocentric distance of the CME leading edge at the time a CME becomes a full halo; this height is significantly shorter in cycle 24 (by ~20%) and has a lower cutoff at ~6 Rs. These results show that cycle-24 CMEs become halos sooner and at a lower speed than the cycle-23 ones. On the other hand, the flare sizes are very similar in the two cycles, ruling out the possibility of eruption characteristics contributing to the differing CME properties. In summary, this study reveals the effect of the reduced total pressure in the heliosphere that allows cycle-24 CMEs expand more and become halos sooner than in cycle 23. Our findings have important implications for the space-weather consequences of CMEs in cycle 25 (predicted to be similar to cycle 24) and for understanding the disparity in halo counts reported by automatic and manual catalogs.

## 1. Introduction

Halo coronal mass ejections (CMEs) are normal CMEs that appear to surround the occulting disk of a coronagraph in sky-plane projection (Howard et al. 1982; 1985). The extended FOV of the Solar and Heliospheric Observatory (SOHO) coronagraphs have shown that halo CMEs are an important subset of CMEs that are fast and wide on the average (Webb et al. 2000; Webb 2002; Gopalswamy et al. 2003; Zhao and Webb 2003; St. Cyr 2005; Gopalswamy et al. 2007, 2010a; Lamy et al. 2019). The fraction of halo CMEs as an indicator of the energy of a CME population (Gopalswamy 2010): 60-70% of CMEs associated with magnetic clouds (MCs), non-MCs, interplanetary Type II radio bursts, interplanetary shocks, intense geomagnetic storms, and large solar energetic particle (SEP) are halos. All CMEs (100%) associated with solar gamma-ray events lasting ≥3 hours are halos (Gopalswamy et al. 2019a). Limb halos are asymmetric halos: eruptions from one limb of the Sun cause disturbances (shocks) above the opposite limb. The expansion of these CMEs must be enormous in that the associated shocks need to have an angular extent >180º. Limb halos are less geoeffective due to the glancing blow they deliver to Earth's magnetosphere (Gopalswamy et al. 2005; 2007), but occasionally they can produce intense geomagnetic storms due to their sheath (Huttunen et al. 2002; Gopalswamy et al. 2010b; Cid et al. 2012).

Comparing the widths of CMEs in cycles 23 and 24 associated with flares originating within 30º from the limb and having flare size ≥C3.0, Gopalswamy et al. (2014a) reported that cycle-24 CMEs are wider, although the speeds are similar (average: 658 km/s in cycle 23 vs. 688 km/s in cycle 24). Comparing all halo CMEs observed by SOHO coronagraphs in cycles 23 and 24 (irrespective of the source location) Gopalswamy et al. (2015a) reported that the halo numbers are similar in the two cycles unlike the sunspot number (SSN). The average halo-CME speeds are not different (933 km/s in cycle 23 vs. 962 km/s in cycle 24), while the cycle-24 halos are widespread in longitude. These findings were attributed to back-reaction of the heliosphere on CME properties: CMEs expand anomalously in cycle 24 due to the reduced heliospheric pressure (Gopalswamy et al. 2014a). One of the implications of the inflated width is that cycle-24 CMEs should become halos at a lower speed and at shorter distances from the Sun. Disk halos are not useful for speed comparison because of the projection effects. Limb halos have minimal projection effects, so they are well suited. We now have large samples of limb halos in two whole cycles, so we can test this prediction (the limb-halo samples were small in Gopalswamy et al. 2014a). One might also wonder if the heliospheric state is solely responsible for the inflated CME widths in cycle 24. We answer this question by comparing the solar source properties, using soft X-ray flare sizes, which are generally large for energetic CMEs such as halos. This work provides further evidence that the heliospheric state determines CME properties using a unique CME population – the limb halos.

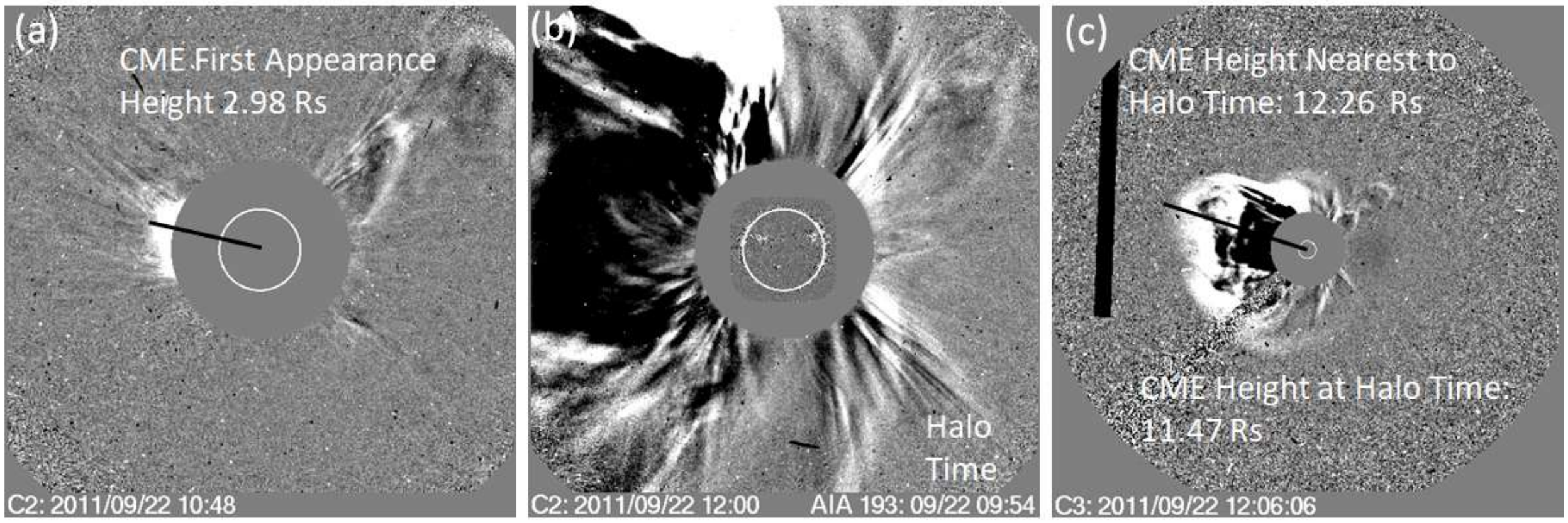

**Figure 1.** The 2011 September 22 east-limb halo at (a) first appearance (10:48 UT) in LASCO/C2 FOV at a height of 2.98 Rs, (b) the "halo time" when the disturbances reach the west limb (11:36 UT). (c) 11:42 UT when the CME height is measured (H=12.26 Rs) and extrapolated to the halo time using local speed (1905 km $s^{-1}$) to get the halo height (11.47 Rs).

## 2. Data

We use limb halos observed by SOHO's Large Angle and Spectrometric Coronagraph (LASCO, Brueckner et al. 1995) available at https://cdaw.gsfc.nasa.gov/ CME_list/halo/halo.html (Gopalswamy et al. 2010a). The catalog lists CME sky-plane and space speeds, eruption location, flare start time, and soft X-ray flare size. We extract limb halos whose sources have central meridian distance (CMD) in the range 60º-90º, although CMD = 90º include some behind the limb events. Backside events are readily identified using the Sun Earth Connection Coronal and Heliospheric Investigation (SECCHI, Howard et al. 2008) suit onboard the Solar Terrestrial Relations Observatory (STEREO) only in cycle 24, so we stay with the single view. For the same reason, we do not consider halos observed by STEREO coronagraphs (Vourlidas et al. 2017).

Figure 1 shows an example: the 2011 September 22 halo, first appearing in the LASCO/C2 FOV at 10:48 UT, erupting from N09E89 in association with an X1.4 flare. The nose is the fastest moving section of the CME above the source regions and always readily identified in LASCO images as indicated in Fig.1. The CME sky-plane and space speeds are the same (1905 km $s^{-1}$) as the source is at the limb. The peak speed is ~2400 km $s^{-1}$ determined from a Graduated Cylindrical Shell (GCS, Thernisien 2011) model fit to SOHO and STEREO images, but the average speed within the coronagraph FOV is close to the sky-plane speed (Gopalswamy et al. 2014b). At 11:36 UT, the CME-associated disturbances appeared on the west limb; this is the halo time (HT), when the CME leading edge (LE) has moved beyond LASCO/C2 FOV. In the 11:42 UT LASCO/C3 image, the LE is at ~12.26 Rs, which extrapolates to 11.47 Rs as the height at HT ($h_{HT}$). It is straight forward to determine $h_{HT}$ because the nose is well within the LASCO/C3 FOV at HT. We followed this procedure to determine $h_{HT}$ for 44 cycle-23 and 38 cycle-24 limb halos (see Table 1).

Table 1 lists the serial number, CME date and first-appearance time in columns 1-3. Columns 4 and 5 give the eruption location and GOES soft X-ray flare size. $h_{HT}$ is given in column 6. On 2010 August 10, the nominal cadence of LASCO C2 increased from 3 to 5 images/hour. One might think $h_{HT}$ became smaller after the cadence change because our cycle-24 events occurred after this date. We computed $h_{HT}$ by reducing the cycle-24 cadence to match that in cycle 23 and given in parentheses (column 6). The sky-plane ($V_{Sky}$) and space speeds ($V_{Sp}$) are given in columns 7 and 8, respectively. Space speeds are expected be not more than 15% higher corresponding to an event originating at ~60º longitude. The data are incomplete (DG) for three cycle-23 halos. On 2003 November 11, a data gap from 13:54 to 15:30 UT prevented the determination of HT and $h_{HT}$. The 2005 January 20 CME was observed only in one LASCO/C2 frame; the subsequent images were corrupted by the intense energetic particle event (e.g., Gopalswamy et al. 2012). Although the halo nature can be discerned from the images, it is difficult to make meaningful measurements. When the 2006 December 6 CME appeared, the leading edge was beyond the LASCO/C2 FOV; there were no LASCO/C3 data, so no measurements are possible. We exclude these three events. Column 9 indicates whether a large SEP event is associated with the limb halos (Y – Yes, N – No, M – minor, and HiB – high background due to previous events). Column 10 gives information on the associated type II radio burst from Wind/WAVES (Bougeret et al. 1995) observations:

(https://cdaw.gsfc.nasa.gov/CME_list/radio/waves_type2.html, Gopalswamy et al. 2019b). If a type II burst is present at decameter-hectometric (DH) wavelengths, the frequency range is noted. “N” indicates the absence of a type II burst. “Nm” indicates that a metric type II burst is associated with the CME, but not a DH type II burst.

## 3. Analysis and Results

In this section we compare the speed and $h_{HT}$ distributions in solar cycles 23 and 24. We compare the cycle-23 $h_{HT}$ with cycle-24 ones obtained using regular and reduced cadences. We perform Kolmogorov-Smirnov (KS) two-sample tests to compare CME and flare properties between the two cycles.

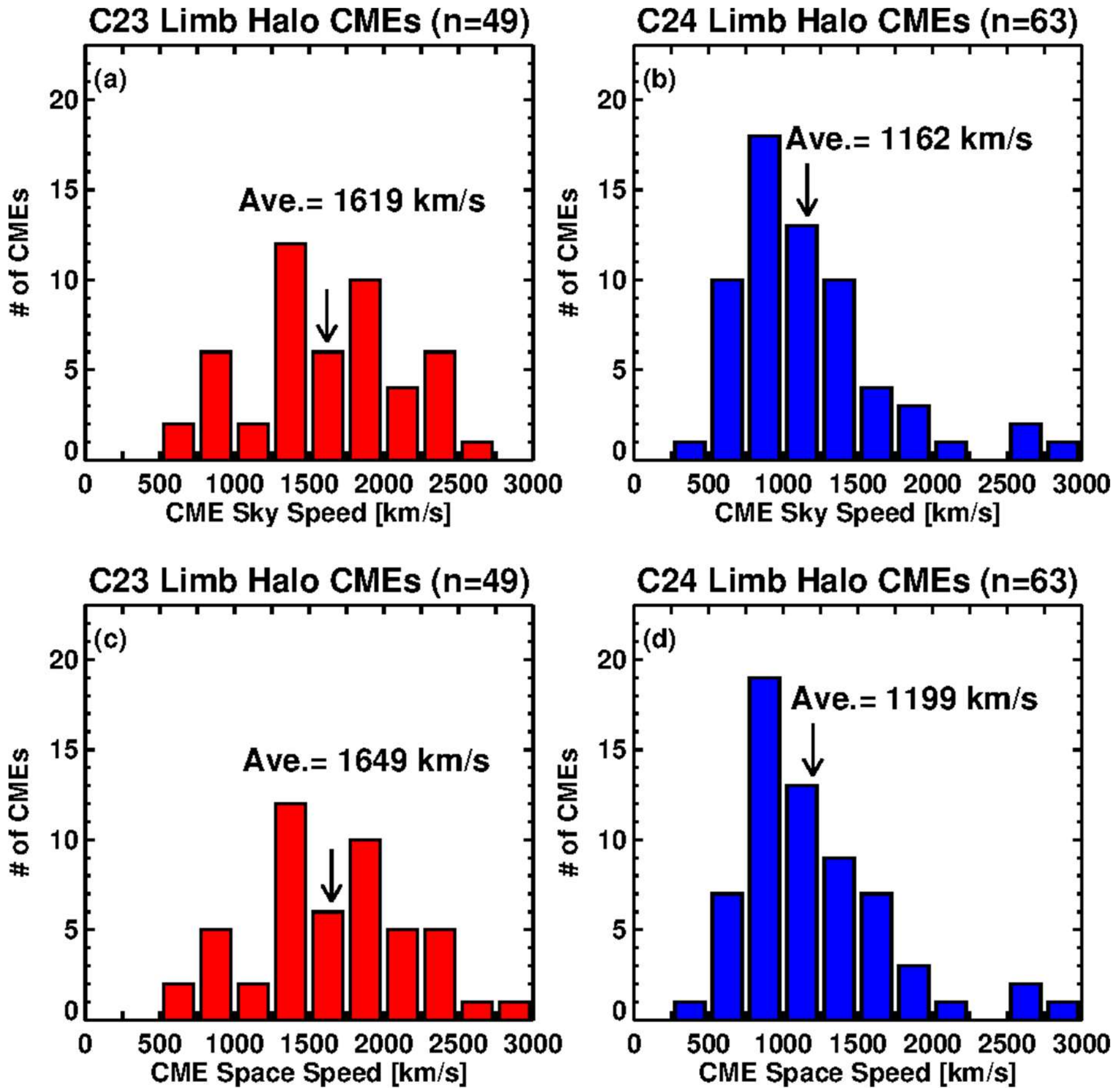

**Figure 2.** The sky-plane (a,b) and space speed (c,d) distributions of 41 cycle-23 and 38 cycle-24 halos. The distribution averages are noted.

### 3.1 CME speeds

Table 1 shows that the cycle-23 sky-plane (space) speeds range from 556 (563) km s$^{-1}$ to 2657 (2662) km s$^{-1}$, with a similar range in cycle 24: 505 (516) km s$^{-1}$ to 3163 (3163) km s$^{-1}$. Even the lowest speeds in the samples are higher than the average speed of the general population of CMEs (~450 km s$^{-1}$, see Gopalswamy 2010). For all limb halos in the two cycles the average sky-plane speed is 1464 ± 129 km s$^{-1}$, which is higher than that of all CMEs by a factor >3. The average sky-plane and space speeds are similar in each cycle, but they are quite different between the two cycles (Fig. 2). The cycle-23 average sky-plane speed is 1637 ± 156 km s$^{-1}$, similar to the average space speed (1655 ± 156 km s$^{-1}$). Similarly, the cycle-24 average sky-plane speed is 1281 ± 202 km s$^{-1}$, not too different from the average space speed (1297 ± 202 km s$^{-1}$). The cycle-24 speed is thus ~28% smaller than that in cycle 23. The cycle-23 speed distribution is normal (median speed: 1594 km s$^{-1}$), while the cycle-24 distribution is lognormal (median speed: 1094 km s$^{-1}$). A KS comparison of the two sky-plane speeds yields a test statistic D = 0.4365 with a corresponding chance coincidence probability p = 0.001, indicating a highly-significant D value. For sample sizes of 41 and 38 in the two cycles, the critical value $D_c$ = 0.3062 at 95% confidence level, confirming that the speed difference between the cycles is significant (D > Dc). The D value is the same when space speeds are used in the test. There was no speed difference between limb CMEs (cycle 23: 658 km s$^{-1}$; cycle 24: 688 km s$^{-1}$) associated with >C3.0 flares although the widths were significantly different (Gopalswamy et al. 2014a).

### 3.2 CME Leading-Edge Height at Halo Time

Since the sky-plane and space speeds are similar, we determine $h_{HT}$ using sky-plane measurements. The cycle-23 $h_{HT}$ ranges from 7.98 Rs to 25.22 Rs (average: 13.33 ± 1.14 Rs); cycle-24 $h_{HT}$ ranges from 6.96 Rs to 19.12 Rs (average: 11.15 ± 0.99) Rs (Fig. 3). The cycle-23 average $h_{HT}$ is larger than that in cycle 24 by ~19.6%. A KS test of the two $h_{HT}$ distributions yields D = 0.3280 with a corresponding p=0.022 indicating a statistically-significant difference (at 95% confidence level, D > $D_c$ = 0.3062). Reducing the cycle-24 image cadence to match that in cycle 23, the cycle-24 $h_{HT}$ shows a small increase (~4%): 11.60 ± 0.97 Rs (Fig. 3c). Slower CMEs and smaller $h_{HT}$ in cycle 24 indicate that halos are formed sooner and at lower speeds in cycle 24, confirming our prediction. The weak state of the heliosphere in cycle 24 allow the CMEs expand more become halos sooner.

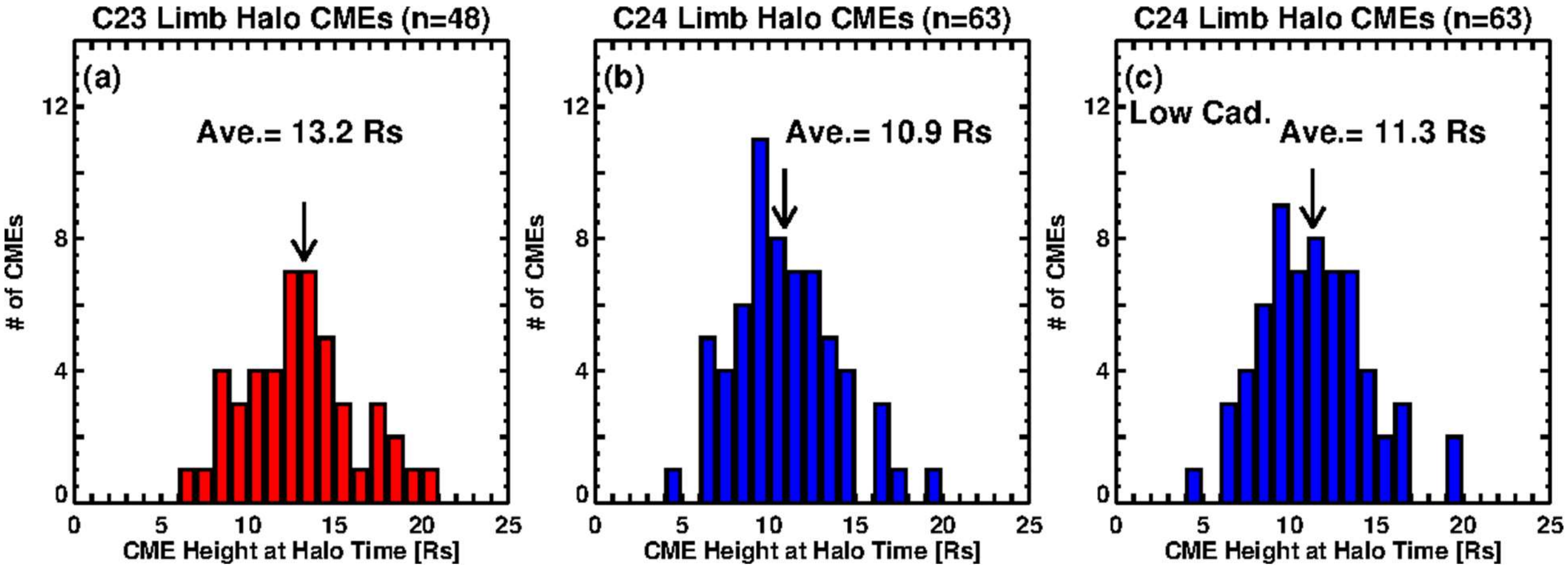

**Figure 3.** Distributions of halo heights (a) in cycle 23, (b,c) in cycle 24 with full and reduced cadences. The averages of the distributions are noted.

Figure 4 shows a weak but significant correlation between the speed and $h_{HT}$. The weakest correlation is in cycle 23 with a correlation coefficient r = 0.27, which is still significant (the Pearson critical correlation coefficient ($r_c$) for a sample size of 41 is 0.26 at 95% confidence level). In cycle 24, the correlation is stronger (r = 0.51 vs. $r_c$ = 0.25). This correlation simply means that with a given image cadence, faster CMEs will be observed in fewer frames within the FOV and therefore likely to be observed at larger heights at HT, which would have been estimated to be earlier if the cadence were higher. There is some indication of this effect shown in Fig. 3. More interesting is the result that the data points in the two cycles are well separated around 1300 km $s^{-1}$, which is close to the average speed of cycle-24 CMEs. The cycle-24 (blue) data points are clustered at the lower left of the plot, while the cycle-23 (red) data points are at the upper right. The clustering suggests that cycle-24 CMEs become halos sooner at lower speeds, while the cycle-23 CMEs take longer and must be faster to become halos. In other words, the cycle-23 CMEs need to work harder against the heliospheric total pressure to expand and become halos.

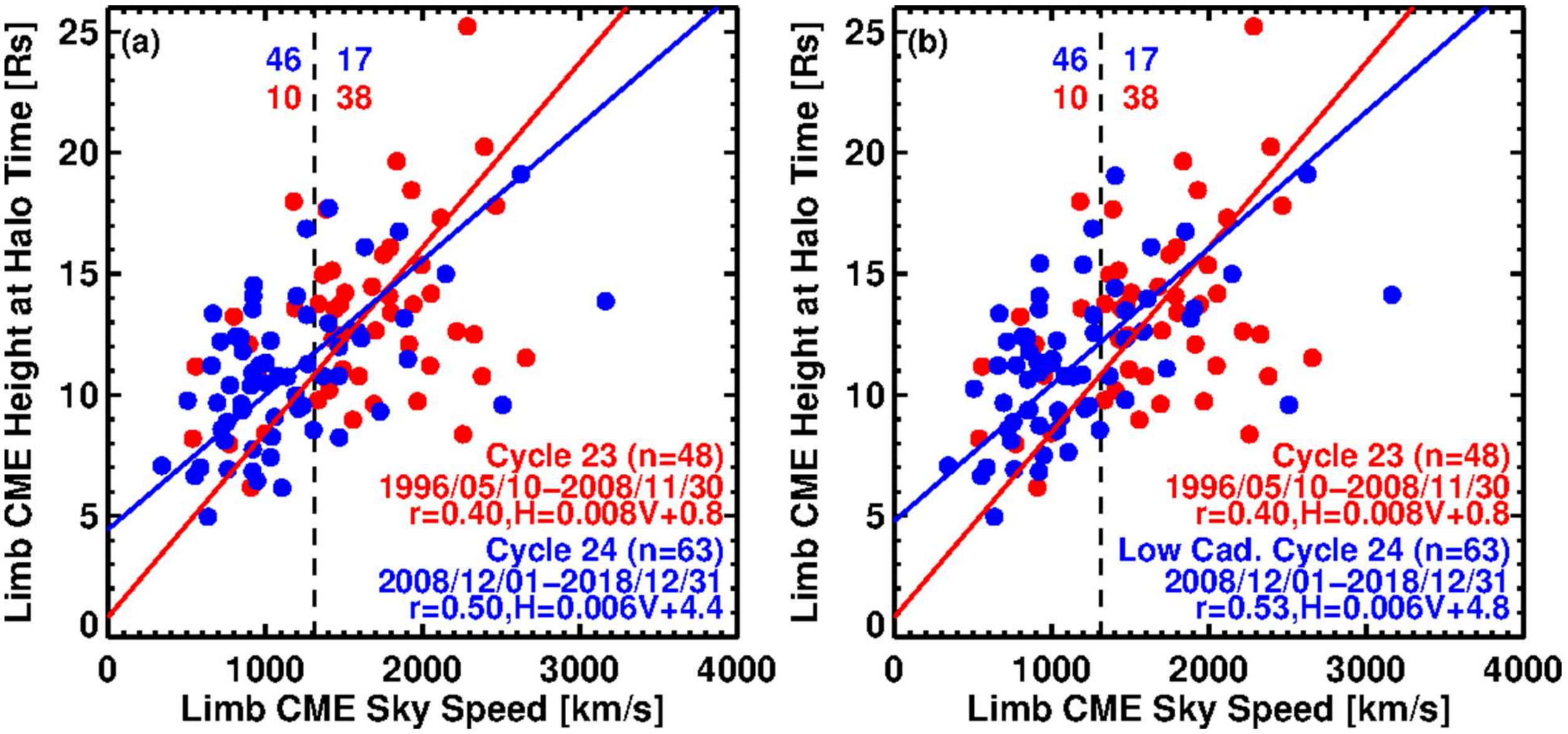

**Figure 4.** Scatter plots of $h_{HT}$ vs. sky-plane speed in cycles 23 (red) and 24 (blue). (a) $h_{HT}$ obtained using the actual cadences, and (b) cycle-24 $h_{HT}$ obtained using reduced cadence. The best-fit lines and their equations are shown in red and blue for cycles 23 and 24, respectively (V: speed, H: halo height). The cycle-23 (red) and cycle-24 (blue) halos cluster on either side of ~1300 km $s^{-1}$. The cycle-24 data points are generally at the lower left, while the cycle-23 data points at the upper right, albeit some overlap.

### 3.3 Flare size comparison

Flare sizes indicate how much of the released energy in an eruption is converted into thermal energy. The cycle-23 flare sizes range from B8.4 to X28 (15 X-, 19 M-, 9 C- and 1 B-class flares) with a median of M4.7 (see Table 1). In cycle 24, the range is from B5.4 to X8.2 (16 X-, 13 M-, 6 C-, and 3 B-class flares) with a median of M6.8. The M- and X-class flares dominate in the two samples with similar fractions: 77% (SC 23) and 76% (SC 24). Evidently, the flare sizes are not significantly different between the two cycles. A KS test of the two sets of flare sizes yields D = 0.1400 with p=0.789 indicating that the two flare size distributions are not significantly different. At 95% confidence level, D < $D_c$ = 0.3012 for sample sizes of 44 (cycles

23) and 38 (cycle 24). Flares and CMEs are manifestations of the same energy release and products of the magnetic reconnection process in the source region. While the flare structure is confined to the Sun and not affected by the heliospheric state, CMEs propagate into the heliosphere and interact with it. Thus, similar flare sizes and differing CME properties are consistent with the differing heliospheric state in the two solar cycles.

### 3.4 Particle acceleration

Large SEP events detected in space are indicative of particle acceleration by CME-driven shocks. While electrons are also identified in space, a better indicator is the presence of type II radio bursts. Table 1 shows that 17 of the 25 (or 68%) cycle-23 western halos and10 of the 17 (or 59%) in cycle 24 have SEP association. Only 3 eastern halos in each cycle have SEP association. In both cycles, there are many minor (M) and high-background (HiB) events. Only one western halo in cycle 23 and 2 in cycle 24 have no SEP association. The non-SEP CME in cycle 23 is from the 1999 July 25 high-latitude (N38W81) eruption. The two non-SEP CMEs in cycle 24 are slow: 505 km $s^{-1}$ (2012 February 23) and 921 km $s^{-1}$ (2012 April 9) compared to the typical speed (~1550 km $s^{-1}$) of CMEs associated with large SEP events (e.g., Gopalswamy 2018).

SEP events magnetically connected to the observer are detected; type II bursts do not have such a requirement. Only 2 of the 44 cycle-23 limb halos are not associated with DH type II bursts; they are slow (800 km $s^{-1}$ on 2000 October 24; 896 km $s^{-1}$ on 2002 January 4) compared to the average speed of CMEs associated with cycle-23 DH type II bursts (1219 km $s^{-1}$). The 2002 January 4 CME is associated with a metric type II burst, while the 2000 October 24 CME is radio-quiet. Nine cycle-24 halos have no DH type II association (see Table 1): five have metric type II bursts, three are quiescent filament-eruption events (which are only occasionally associated with DH type II bursts, Gopalswamy et al. 2015b), and the slowest cycle-24 halo (2012 February 23). The four radio-quiet halos are slow (505 to 919 km $s^{-1}$) compared CMEs associated with cyle-24 DH type II bursts (1059 km $s^{-1}$). Thus, an overwhelming majority of limb halos (cycle 23: 98%; cycle 24: 90%) have type II bursts indicative of electron acceleration.

## 4. Discussion and Summary

We analyzed large samples of limb halo CMEs observed in cycles 23 and 24. We found that cycle-24 halos are slower than the cycle-23 ones. The limb CMEs that revealed wider width in cycle 24 did not show the speed difference (Gopalswamy et al. 2014a). We introduced a new parameter – the height at halo time is readily determined for limb halos, but not for disk halos (the CME nose is hidden by the occulting disk of the coronagraph). The cycle-24 halo heights are significantly smaller. Combined with the result that cycle-24 halos are slower, we conclude that halos form at shorter heliocentric distances at lower speeds. Furthermore, we ruled out the possibility that the difference in CME properties is due to solar-source characteristics represented by soft X-ray flare size. Thus, we can pin down the heliospheric state as the main cause of the anomalous CME expansion in cycle 24.

The number of limb halos (44) in cycle 23 is only slightly larger than the number (38) in cycle 24. This corresponds to a drop of only 14% in cycle 24. There was a 4-month SOHO data gap in cycle 23 (3 months in 1998 and one month in 1999). Assuming that the limb halos occurred at the cycle-averaged monthly rate (0.3 per month), expect only one additional CME during the data gap. Then the drop is only 16%. This is much smaller than the 40% drop in the cycle-averaged total SSN from 81 to 49. Normalizing to SSN, we see that there are ~42% more limb halos per SSN in cycle 24. This result was previously obtained by considering all LASCO halos in the first half of the two cycles (Gopalswamy et al. 2015a).

The effect of the heliospheric state on CMEs has important implications for space-weather: the lower geoeffectiveness in cycle 24 results from the lower CME speed and weaker magnetic field (due to expansion). Furthermore, milder space weather is expected in cycle 25, which has been predicted to be similar to cycle 24. This study also helps understand the disparity between manual and automatic catalogs in identifying halos. For example, the ARTEMIS catalog reported only 11 halos in cycle 23 (Lamy et al. 2019), while the SOHO/LASCO manual catalog reported nearly 400 halos. The lower cutoff of ~6 Rs in the halo-height distribution (Fig. 3) is very close to the outer edge of LASCO/C2 FOV. This might explain why automatic catalogs that use LASCO/C2 data, report very few halos. Our limb-halo data set will serve as a reference and ground truth to evaluate the success/failure of automatic catalogs.

The main results of this paper can be summarized as follows:

1. Limb halos are one of the fastest of CME populations with an average speed of 1464 ± 129 km $s^{-1}$, with a high degree of SEP-event and type II-burst association and are mostly associated with M- and X-class flares.

2. The cycle-23 limb halos are significantly faster (1637 ± 164 km $s^{-1}$) than the cycle-24 ones (1281 ± 202 km $s^{-1}$).

3. A new parameter – the halo height characterizes the influence of the heliospheric state on CMEs. The average cycle-23 halo height (13.33 ± 1.14 Rs) is significantly larger than the cycle-24 value (11.15 ± 0.99 Rs).

4. The speed and halo-height differences indicate that cycle-24 CMEs become halos sooner at lower speeds, consistent with the effect of weak heliospheric state. In the initial study that reported wider CMEs in cycle 24 (Gopalswamy et al. 2014a), the limb CMEs had similar speeds.

5. The flare-size distributions in the two cycles are similar with median values of M4.7 (cycle 23) and M6.8 (cycles 24), indicating that the heliospheric state rather than the solar-source properties is responsible for the differing CME properties.

6. While there is a high degree of association between limb halos and shocks, the reduced association in cycle 24 is consistent with the reduced efficiency of particle acceleration.

**Acknowledgments.** This work benefitted from NASA's open data policy in using SOHO, STEREO, and SDO data and NOAA's GOES X-ray data. SOHO is a joint project of ESA and NASA. STEREO is a mission in NASA's Solar Terrestrial Probes program. Work supported by NASA's LWS TR&T and heliophysics GI programs.

Table 1. List of Limb Halo CMEs from Solar Cycles 23 and 24

| # | CME Date [UT] | Time [UT] | Location | Flare Size | Halo Height [Rs] | $V_{Sky}$ [km/s] | $V_{Sp}$ [km/s] | SEP[e] | Type II[f] [MHz] |
|---|---|---|---|---|---|---|---|---|---|
| | | | | Cycle 23 | | | | | |
| 1 | 1997/11/06 | 12:10:41 | S18W63 | X9.4 | 8.97 | 1556 | 1604 | Y | 14 – 0.1 |
| 2 | 1998/04/23 | 05:55:22 | S17E106 | X1.4 | 9.62 | 1691 | 1691 | HiB | 14 – 0.2 |
| 3 | 1998/11/24 | 02:30:05 | S28W103 | X1.0 | 13.40 | 1798 | 1798 | M | 1 – 0.4 |
| 4 | 1999/07/25 | 13:31:21 | N38W81 | M2.4 | 17.67 | 1389 | 1392 | N | 0.2? |
| 5 | 2000/04/04 | 16:32:37 | N17W60 | C9.7 | 13.57 | 1188 | 1372 | Y | 14 – 0.2 |
| 6 | 2000/05/05 | 15:50:05 | S17W100 | M1.5 | 10.77 | 1594 | 1594 | M | 14 – 2.5 |
| 7 | 2000/10/16 | 07:27:21 | N03W108 | M2.5 | 9.78 | 1336 | 1336 | Y | 14 – 1 |
| 8 | 2000/10/24 | 08:26:05 | S23E70 | C2.3 | 13.24 | 800 | 820 | N | N |
| 9 | 2000/10/25 | 08:26:05 | N09W63 | C4.0 | 7.98 | 770 | 813 | Y | 10 – 0.3 |
| 10 | 2001/04/01 | 11:26:06 | S22E108 | M5.5 | 13.72 | 1475 | 1475 | HiB | DG |
| 11 | 2001/08/19 | 06:06:05 | N30W75 | B8.4[a] | 11.18 | 556 | 563 | HiB | 1 – 0.4 |
| 12 | 2001/10/01 | 05:30:05 | S24W81 | M9.1 | 10.19 | 1405 | 1409 | Y | 1 – 0.15 |
| 13 | 2001/11/22 | 20:30:33 | S25W67 | M3.8 | 13.55 | 1443 | 1472 | Y | 8 - 1 |
| 14 | 2001/12/14 | 09:06:06 | N07E97 | M3.5 | 14.22 | 1506 | 1507 | N | 0.7 – 0.3 |
| 15 | 2001/12/28 | 20:30:05 | S24E104 | X3.4 | 12.63 | 2216 | 2216 | Y | 14 – 0.35 |
| 16 | 2002/01/04 | 09:30:05 | N38E87 | C3.7 | 12.10 | 896 | 907 | HiB | Nm |
| 17 | 2002/01/14 | 05:35:07 | S28W108 | M4.4 | 11.05 | 1492 | 1492 | Y | 14 – 0.35 |
| 18 | 2002/02/20 | 06:30:05 | N12W72 | M5.1 | 10.80 | 952 | 965 | Y | 14-10? |
| 19 | 2002/03/10 | 23:06:55 | S22E113 | M2.3 | 12.31 | 1429 | 1429 | N | 14 – 8 |
| 20 | 2002/03/22 | 11:06:05 | S10W90 | M1.6 | 15.79 | 1750 | 1750 | Y | 14 – 0.5 |
| 21 | 2002/04/21 | 01:27:20 | S14W84 | X1.5 | 20.25 | 2393 | 2396 | Y | 10 – 0.06 |
| 22 | 2002/07/19 | 16:30:05 | S15E115 | C2.9[b] | 11.20 | 2047 | 2047 | HiB | 5 – 1 |
| 23 | 2002/07/20 | 22:06:09 | S13E99 | X3.3 | 13.74 | 1941 | 1941 | M | 10 – 2 |
| 24 | 2002/07/23 | 00:42:05 | S13E72 | X4.8 | 25.22[c] | 2285 | 2318 | HiB | 11 – 0.4 |
| 25 | 2002/08/22 | 02:06:06 | S07W62 | M5.4 | 8.42 | 998 | 1034 | Y | 14 – 3.5 |
| 26 | 2002/08/24 | 01:27:19 | S02W81 | X3.1 | 12.09 | 1913 | 1920 | Y | 5 – 0.4 |
| 27 | 2002/12/08 | 23:54:05 | S18E70 | C2.5 | 13.76 | 1339 | 1361 | N | DG |
| 28 | 2003/05/31 | 02:30:19 | S07W65 | M9.3 | 19.65[c] | 1835 | 1888 | Y | 3 – 0.15 |
| 29 | 2003/06/15 | 23:54:05 | S07E80 | X1.3 | 14.18 | 2053 | 2062 | N | 14 – 0.4 |
| 30 | 2003/11/04 | 19:54:05 | S19W83 | X28. | 11.53 | 2657 | 2662 | Y | 10 – 0.2 |
| 31 | 2003/11/11 | 13:54:05 | S03W61 | M1.6 | DG[d] | 1315 | 1367 | HiB | 1 – 0.5? |
| 32 | 2004/07/29 | 12:06:05 | N00W90 | C2.1 | 18.00 | 1180 | 1180 | M | 1 – 0.05 |
| 33 | 2005/01/20 | 06:54:05 | N14W61 | X7.1 | DG[d] | ---- | ---- | Y | 14 –0.025 |
| 34 | 2005/06/03 | 12:32:10 | N15E97 | M1.0 | 14.48 | 1679 | 1679 | N | 10 – 0.27 |
| 35 | 2005/07/13 | 14:30:05 | N11W90 | M5.0 | 15.14 | 1423 | 1423 | M | 14 – 1 |
| 36 | 2005/07/14 | 10:54:05 | N11W90 | X1.2 | 17.31 | 2115 | 2115 | Y | 3 – 0.8 |
| 37 | 2005/07/27 | 04:54:05 | N11E104 | M3.7 | 14.08 | 1787 | 1787 | Y | 1 – 0.45 |
| 38 | 2005/07/30 | 06:50:28 | N12E60 | X1.3 | 9.73 | 1968 | 2043 | Y | 9 – 0.08 |
| 39 | 2005/08/22 | 17:30:05 | S13W65 | M5.6 | 10.77 | 2378 | 2445 | Y | 12 – 0.04 |
| 40 | 2005/08/23 | 14:54:05 | S12W70 | M2.7 | 18.45 | 1929 | 1929 | HiB | 13 – 0.2 |
| 41 | 2005/09/05 | 09:48:05 | S07E119 | C2.7[b] | 12.50 | 2326 | 2334 | N | 1.5 – 0.06 |
| 42 | 2005/09/09 | 19:48:05 | S12E67 | X6.2 | 8.37 | 2257 | 2311 | HiB | 14 – 0.05 |
| 43 | 2006/12/06 | 20:12:05 | S05E64 | X6.5 | DG[d] | ---- | ---- | Y | 14 – 0.03 |
| 44 | 2007/01/25 | 06:54:04 | S08E102 | C6.3 | 14.97 | 1367 | 1367 | N | 14 – 0.09 |
| | | | Cycle 24 | | | | | | |
| 1 | 2011/08/09 | 08:12:06 | N17W69 | X6.9 | 12.34 (13.98) | 1610 | 1640 | Y | 14 – 0.2 |
| 2 | 2011/09/22 | 10:48:06 | N09E89 | X1.4 | 11.47 (13.57) | 1905 | 1905 | Y | 14 – 0.07 |
| 3 | 2011/10/22 | 10:24:05 | N25W77 | M1.3 | 10.48 (11.49) | 1005 | 1011 | Y | 1.5 – 0.1 |
| 4 | 2012/01/16 | 03:12:10 | N34E86 | C6.5 | 9.08 (9.08) | 1060 | 1060 | N | 3 – 0.9 |
| 5 | 2012/01/26 | 04:36:05 | N41W84 | C6.4 | 9.98(10.84) | 1194 | 1195 | HiB | 5 – 0.5? |
| 6 | 2012/01/27 | 18:27:52 | N27W78 | X1.7 | 9.58 (9.58) | 2508 | 2541 | Y | 14 – 0.15 |
| 7 | 2012/02/09 | 21:17:36 | N18E80 | B8.0[b] | 11.21 (11.21) | 659 | 663 | N | N[g] |
| 8 | 2012/02/23 | 08:12:06 | N27W71 | B5.4 | 9.77 (10.26) | 505 | 516 | N | N |
| 9 | 2012/03/04 | 11:00:07 | N19E61 | M2.0 | 8.55 (8.55) | 1306 | 1352 | M | 1 – 0.2 |
| 10 | 2012/03/13 | 17:36:05 | N17W66 | M7.9 | 13.16 (13.16) | 1884 | 1931 | Y | 14 – 0.2 |
| 11 | 2012/04/09 | 12:36:07 | N20W65 | C3.9 | 7.75 (8.74) | 921 | 945 | N | 14 - 5 |
| 12 | 2012/05/17 | 01:48:05 | N11W76 | M5.1 | 12.63 (12.63) | 1582 | 1596 | Y | 14 – 0.3 |
| 13 | 2012/07/19 | 05:24:05 | S13W88 | M7.7 | 16.11 (16.11) | 1631 | 1631 | Y | 5 – 0.6 |
| 14 | 2012/11/08 | 02:36:06 | N13E89 | M1.7 | 11.81 (11.81) | 855 | 855 | M | Nm |
| 15 | 2012/11/27 | 02:36:05 | N13E68 | C6.0[b] | 12.38 (12.38) | 844 | 874 | N | N[g] |
| 16 | 2013/05/13 | 02:00:05 | N11E90 | X1.7 | 11.28 (12.55) | 1270 | 1270 | N | 14 - 2 |
| 17 | 2013/05/13 | 16:07:55 | N11E85 | X2.8 | 16.75 (16.75) | 1850 | 1852 | M | 14 -0.3 |
| 18 | 2013/05/14 | 01:25:51 | N08E77 | X3.2 | 19.12 (19.12) | 2625 | 2645 | M | 14 – 0.24 |
| 19 | 2013/05/15 | 01:48:05 | N12E64 | X1.2 | 10.78 (10.78) | 1366 | 1408 | Y | 14 – 0.3 |
| 20 | 2013/05/22 | 13:25:50 | N15W70 | M5.0 | 10.78 (12.32) | 1466 | 1491 | Y | 14 – 0.15 |
| 21 | 2013/09/24 | 20:36:05 | N26E70 | B6.5 | 10.92 (10.92) | 919 | 932 | N | N[g] |
| 22 | 2013/10/25 | 08:12:05 | S08E73 | X1.7 | 7.01 (7.01) | 587 | 599 | N | Nm |
| 23 | 2013/10/25 | 15:12:09 | S06E69 | X2.1 | 10.79 (10.79) | 1081 | 1103 | M | 14 – 0.2 |
| 24 | 2013/10/28 | 02:24:05 | N04W66 | X1.0 | 9.67 (9.67) | 695 | 726 | M | Nm |
| 25 | 2013/10/29 | 22:00:06 | N05W89 | X2.3 | 11.33 (11.33) | 1001 | 1001 | M | Nm |
| 26 | 2013/11/19 | 10:36:05 | S14W70 | X1.0 | 8.12 (8.12) | 740 | 761 | M | 14 – 5 |
| 27 | 2014/01/20 | 22:00:05 | S07E67 | C3.6 | 8.58 (8.58) | 721 | 750 | N | 14 – 8 |
| 28 | 2014/02/20 | 08:00:07 | S15W73 | M3.0 | 6.45 (7.50) | 948 | 960 | Y | 12 – 7.7 |
| 29 | 2014/02/25 | 01:25:50 | S12E82 | X4.9 | 14.99 (14.99) | 2147 | 2153 | Y | 14 – 0.1 |
| 30 | 2014/06/10 | 13:30:23 | S17E82 | X1.5 | 11.99 (13.48) | 1469 | 1473 | N | 14 – 9 |
| 31 | 2014/08/24 | 12:36:05 | S07E75 | M5.9 | 6.65 (6.65) | 551 | 569 | N | Nm |
| 32 | 2015/02/09 | 23:24:05 | N12E61 | M2.4 | 6.16 (7.63) | 1106 | 1148 | N | 14 – 9? |
| 33 | 2015/03/07 | 22:12:05 | S19E74 | M9.2 | 16.87 (16.87) | 1261 | 1304 | N | 14 – 8? |
| 34 | 2015/04/23 | 09:36:05 | N12W89 | M1.1 | 9.39 (9.39) | 857 | 864 | M | 3 – 1 |
| 35 | 2015/05/05 | 22:24:05 | N15E79 | X2.7 | 12.21 (12.21) | 715 | 721 | N | 2 – 0.5 |
| 36 | 2016/01/01 | 23:24:04 | S25W82 | M2.3 | 9.31 (11.08) | 1730 | 1734 | Y | 1.1 – 0.3 |
| 37 | 2017/04/18 | 19:48:05 | N14E77 | C5.5 | 14.54 (15.43) | 926 | 932 | N | 2 – 0.5? |
| 38 | 2017/09/10 | 16:00:05 | S09W90 | X8.2[b] | 13.87 (14.13) | 3163 | 3163 | Y | 14 – 0.15 |

[a]Flare size estimated from GOES light curve; [b]The flare is partly occulted; the flare size is estimated from GOES data; [c]Some uncertainty in the halo height due to lower cadence; [d]Insufficient data to determine the height at halo time (DG: data gap); [e]Large SEP event (>10 MeV proton intensity ≥10 particle flux unit, pfu; 1 pfu = 1 particle per $cm^{-2}s^{-1}sr^{-1}$): Y-yes, N – No, M – minor event (proton intensity <10 pfu), HiB – high background due to previous events; [f]Type II burst: the frequency range in DH domain, N – no DH Type II, m – metric type II. [g]Associated with a quiescent filament eruption.